# Stiffness modeling for near field acoustic levitation bearings

Yaoke Wang[1] and Ping Guo[1,*]

[1]*Department of Mechanical Engineering, Northwestern University, Evanston, IL 60208, USA*

*\* Corresponding author: ping.guo@northwestern.edu*

**Abstract**

The dynamic characteristics of near-field levitation bearings have been investigated in this study. Through theoretical analysis, two different types of system stiffness are defined and derived analytically. The dynamic stiffness relates the excitation amplitude to the dynamic force amplitude, while the effective stiffness governs the time-averaged force-displacement relationship. The results indicate two non-linear and asymmetric spring constants that can effectively predict levitation force and height. The models are verified with a carefully designed experimental setup to eliminate the structural resonance effect. Besides, some unique dynamic behaviors are investigated and predicted based on the proposed stiffness model.

Non-contact transmission, such as air and magnetic bearings, has been widely adopted for their high-speed operation, low heat generation, low friction, zero backlashes, etc. The application of these bearings, however, is still limited by their high cost, complex system design, undesired noise, and especially the difficulty in miniaturization.

An alternative non-contact bearing technology has been proposed and advanced based on near-field acoustic levitation.[1] When two very closely placed parallel surfaces (in the range from tens to several hundred microns) are cyclically squeezing the thin air film in between at a high frequency (from several thousand Hertz to an ultrasonic frequency), the time-averaged reaction force from the air medium will be non-zero due to the increased viscosity at a smaller air gap. Therefore, the symmetric and periodic vibration will produce an asymmetrical reaction force, resulting in a substantial levitation force to balance the gravity to achieve non-contact support. This effect is referred to as near field acoustic levitation (NFAL), which has led to alternative designs of non-contact bearings or motors.[2]

Based on near field acoustic levitation (NFAL), linear transportation has been achieved by integrating traveling waves generated by flexural vibration.[3,4] If the levitated object itself is the vibration source, self-levitation can be achieved with unconstrained two-dimensional motion.[5] When the two squeezing surfaces are designed to be curved or circular, non-contact rotary bearings can be realized to support both radial and axial loads.[6] Besides, by coupling the longitudinal and bending modes, the rotary bearing can be transformed from a passive bearing to an active motor by generating bi-directional rotation.[7]

Dynamic behaviors of NFAL bearings are of particular importance and interest for the industrial application of the technology but have not been well studied. Notably, Minikes et al.[8] derived an analytical expression of levitation force based on a perturbation solution to the Reynolds equations. Li et al.[9] discussed the gas inertia and edge effect on NFAL. Melikhov et al. investigated the influences of transition visco-acoustic domain.[10] However, these works were focused on the levitation mechanism and pressure distribution rather than system dynamics. Some rare studies were presented by Bucher et al., [11,12] where the slow and coupled dynamics of a levitated mass were considered.

There has not been a successful attempt to give an analytical expression of bearing stiffness for NFAL. The lack of previous research on dynamic modeling is mainly due to the complex and coupled hydrodynamic equations, which cannot be easily decoupled or analytically solved. Besides, the system behavior is highly nonlinear, which is very challenging in the experimental design for model verification. However, in the prototype design of NFAL devices, an analytical solution that can intuitively approximate the quasi-static and dynamic performance of NFAL will provide tremendous benefits in future research and technical innovation in NFAL. Against this background, we propose a mass-spring model to characterize the system dynamics with two unique system stiffness variables.

The NFAL system can be represented by a spring-mass model as illustrated in Fig. 1(a). The bottom circular surface provides a prescribed vibratory excitation, while the top surface of the same area with a mass $m$ is supported by a virtual spring resulted from the air hydrodynamics. The overall air film thickness can be represented by the summation of an average levitation height and the relative vibration of two parallel surfaces:

$$h(t) = \bar{h}(t) - \tilde{h}(t). \tag{1}$$

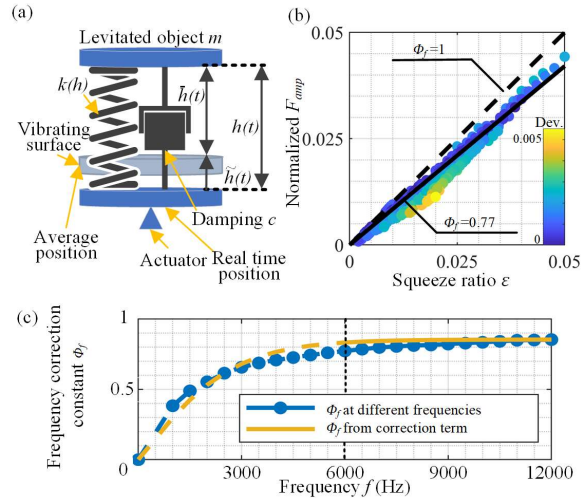

**FIG. 1.** (a) Schematic of the stiffness model; (b) linear approximation of the dynamic force amplitude at 9 kHz; and (c) frequency correction constant $\Phi_f$.

The overbar indicates a time-averaged value, while the tilde symbol indicates a harmonic vibration with its average value equals to zero. Here we assume that the high-frequency variation, $\tilde{h}(t)$, is purely due to the lower surface vibration (actuation), while the levitated object does not oscillate. This assumption is fairly accurate at a high vibration frequency or large levitation mass. In our experiments, with a 1.35-gram levitated object, the responsive vibration from the levitated object will be less than 10% of the actuation amplitude when excited above 6 kHz.

The unique dynamic behaviors of NFAL result in two stiffness variables that characterize the system. When $\tilde{h} \ll \bar{h}$, the responsive force experienced by the levitated object can also be separated into a time-averaged term and a dynamic term as:

$$F(t) = \bar{F} + \tilde{F}_{amp}(t) = \int_{\infty}^{\bar{h}} K(\bar{h})d\bar{h} + k(\bar{h})\tilde{h}(t). \tag{2}$$

The dynamic stiffness, $k$, is valid for the harmonic excitation at a high frequency with a very small squeeze ratio $\varepsilon$, which is defined as the ratio of vibration amplitude to the average air film thickness $A_{mp}/\bar{h}$. The dynamic stiffness governs the instantaneous force response to the levitated object, whereas the effective stiffness, $K$, describes the time-averaged displacement-force response, which is the change of time-averaged levitation $\bar{F}$ due to any additional low-frequency or quasi-static modulation motion added to the existing high-frequency vibration. The effective stiffness resembles more of the conventional definition of bearing stiffness, which relates external force to air gap change; however, it depends on the existence of high-frequency vibration.

The derivations of $k$ and $K$ are given as follows. The pressure distribution in NFAL for a circular surface can be derived by solving the compressible Reynolds equations in polar coordinates:[13]

$$\frac{\partial}{\partial r}\left(prh^3 \frac{\partial p}{\partial r}\right) = 12\eta r \frac{\partial(ph)}{\partial t},$$
$$h = \bar{h} - A_{mp}\sin(2\pi f t), \tag{3}$$
$$B.C.\ p(R,t) = p_a;\ p(r,0) = p_a;\ p'(0,t) = 0,$$

where $r$ is the radial coordinate; $p(r,t)$ describes the pressure distribution at location $r$ and time instance $t$. Other physical constants include the atmospheric pressure $p_a$, time-averaged levitation height $\bar{h}$, levitated object radius $R$, and dynamic viscosity of air $\eta$. Following the definition in Eq. (1), the air film thickness is perturbed by a small harmonic vibration characterized by the vibration amplitude $A_{mp}$ and frequency $f$.

The pressure distribution $p(r,t)$ is calculated using the finite difference method and integrated over the whole surface area to get the reaction force:

$$F(t) = \int_0^R 2\pi p(r,t)r\,dr \approx \bar{F} + \tilde{F}_{amp} e^{i(2\pi ft + \phi)}. \tag{4}$$

We calculated the dynamic force amplitude defined in Eq. (4) under a wide set of process conditions. These conditions are specified in Table 1, which are set according to the achievable range in our proposed experimental setup. We find the dynamic force is independently governed by the squeeze ratio $\varepsilon = A_{mp}/\bar{h}$ and excitation frequency. If we compare the squeeze ratio to the normalized dynamic force, it follows a linear relationship with a constant slope $\Phi_f$ for a given particular excitation frequency as shown in Fig. 1(b), where the local deviations are indicated by the color scale. This linear relationship holds true when the small-amplitude vibration assumption is valid ($\varepsilon < 0.05$). The excitation frequency will influence the slope value $\Phi_f$, which can be approximated by a hyperbolic tangent function shown in Fig. 1(c). For most of the working conditions of NFAL (high-frequency assumption above 6 kHz with a similar or larger object surface area), the frequency correction term $\Phi_f$ can be regarded as a constant. For the simulated conditions listed in Table 1, $\Phi_f$ is determined to be 0.77 at 9 kHz. For the general operating conditions of NFAL devices especially at an ultrasonic frequency, $\Phi_f$ can be simply assumed to be a constant between 0.8 to 1. Please see the Supplementary Material for further discussion on the physical meaning of the frequency correction factor $\Phi_f$.

Table 1 Simulation parameters.

| Frequency $f$ (kHz) | Viscosity $\eta$ (Pa·s) | Object radius $R$ (mm) |
|---|---|---|
| 0.5, 1, 1.5…, 12 | 1.81×10$^{-5}$ | 12.7 |
| Atmospheric pressure (Pa) | Vibration amplitude $A_{mp}$ (μm) | Levitation height $\bar{h}$ (μm) |
| 1.01×10$^5$ | 0.1, 0.2, …, 1 | 20, 21, …, 50 |

By linearizing the solutions to the Reynolds equation using the derived frequency correction constant $\Phi_f$, the force-vibration amplitude relationship and the corresponding dynamic stiffness can be derived as:

$$\tilde{F}_{amp} = \pi R^2 p_a \Phi_f \varepsilon = \frac{\pi R^2 p_a \Phi_f}{\bar{h}} A_{mp},$$

$$k(h) = \frac{\pi R^2 p_a \Phi_f}{h}.$$

(5)

From the results presented in Eq. (5), the dynamic stiffness is inversely proportional to the levitation height. It can be approximated as a non-linear and asymmetric air spring. The variation of dynamic stiffness $k$ is plotted in Fig. 2(a) for the vibration amplitude of 1 μm and the excitation frequency at 9 kHz. The spring will be hardened with the squeeze motion and softened with the release motion, thus resulting in a net positive time-averaged levitation force, as illustrated by Fig. 2(b).

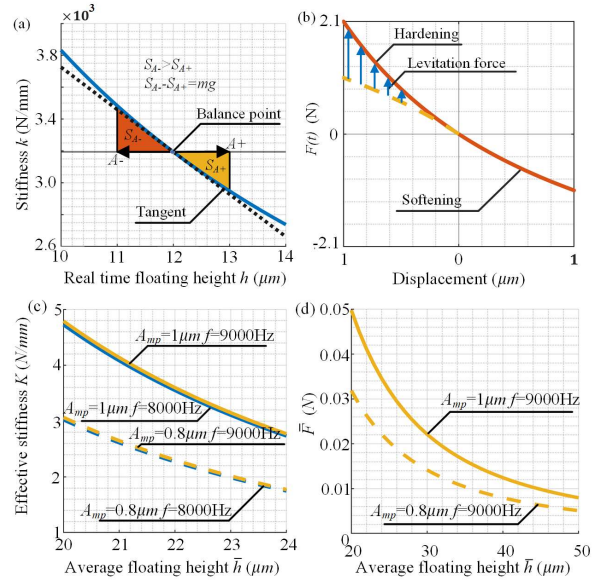

**FIG. 2.** Calculated (a) dynamic stiffness $k$ at 9 kHz and $A_{mp}$ =1 μm (b) force-displacement curves at 9 kHz and $A_{mp}$=1 μm. (c) effective stiffness $K$ and (d) levitation force-height curves.

The change of levitation force concerning the varying average levitation height is determined by another effective stiffness $K$, which governs the low-frequency performance of NFAL bearings (or the behaviors of modulation motion).

By integrating the dynamic force over a vibration cycle, the effective stiffness can be derived as

$$K(\bar{h}) = \frac{\Delta \int_0^{\frac{1}{f}} fk(h)A_{mp}e^{i2\pi ft}dt}{\Delta \bar{h}} = \frac{\Delta \int_0^{\frac{1}{f}} \pi R^2 p_a \Phi_f \varepsilon e^{i2\pi ft} \sum_{n=0}^{\infty}(\varepsilon e^{i2\pi ft})^n}{\Delta \bar{h}} \quad (6)$$

$$\approx \frac{\pi R^2 p_a \Phi_f A_{mp}^2}{\bar{h}^3} = \left(\frac{A_{mp}}{\bar{h}}\right)^2 \frac{\pi R^2 p_a \Phi_f}{\bar{h}} = \varepsilon^2 k(\bar{h}).$$

The change of levitation force due to a displacement input then can be derived as

$$\bar{F} = \int_{\infty}^{\bar{h}} K(\bar{h})d\bar{h} = \frac{\pi R^2 p_a \Phi_f A_{mp}^2}{2\bar{h}^2}, \quad (7)$$

where the integral from $\infty$ to $\bar{h}$ indicates the loading from infinity (no force) to the current levitation height $\bar{h}$.

The effective stiffness $K$ is proportional to the square of the excitation vibration amplitude and inversely proportional to the cubic of the average levitation height. The calculated stiffness curves under two different vibration amplitudes and excitation frequencies are given in Fig. 2(c). It also verifies the assumption that the stiffness is frequency insensitive at high excitation frequencies. The characteristics of levitation force are plotted in Fig. 2(d), indicating an inverse square $\bar{h}$-$\bar{F}$ relationship. When comparing to the derived result of dynamic stiffness $k$, the effective stiffness $K$ is scaled by the square of squeeze ratio $\varepsilon$. Considering the small squeeze ratio assumption, the effective stiffness is 2-4 orders of magnitude smaller than the high-frequency stiffness. Though the presented effective stiffness $K$ is small, our setup and simulated conditions are not designed to achieve maximum efficiency. If the same analysis is applied to the design of Li et al.[6], the calculated effective stiffness $K$ can reach up to 6 N/μm scaled with the surface area and required input power. This result is comparable to a commercial air bearing (LRAP100, *Specialty Components*, USA).

To verify the model accuracy and to obtain reference results for comparison, an experimental setup is built as shown in Fig. 3(a). The actuator is designed to have a much higher resonant frequency beyond the test frequency range, so the structural resonance can be avoided. In addition, it has been validated that the vibration amplitude follows a linear relationship with the input voltage amplitude from 0 to 12 kHz, while the center and edge have the same measured vibration amplitude (no resonant mode). The levitated object is a diamond-turned aluminum disk with a diameter of 25.4 mm, a thickness of 1 mm, and a surface roughness smaller than 0.02μm (*Ra*). To limit the horizontal position variation of the levitated object, three magnetic rods are placed around the actuator. During the experiments, a sinusoidal signal is generated by a DAQ card (PXIe-6366, *NI*, USA), which is amplified by an amplifier (PX200, *Piezodrive*, AUS) to drive the actuator. The displacement of levitated object, which contains the information of levitation height and vibration amplitude, is measured using a laser doppler vibrometer (CLV-2534, *Polytec*, USA), which is recorded by the same DAQ card at a sampling frequency of 1 MHz as shown in Fig. 3(b).

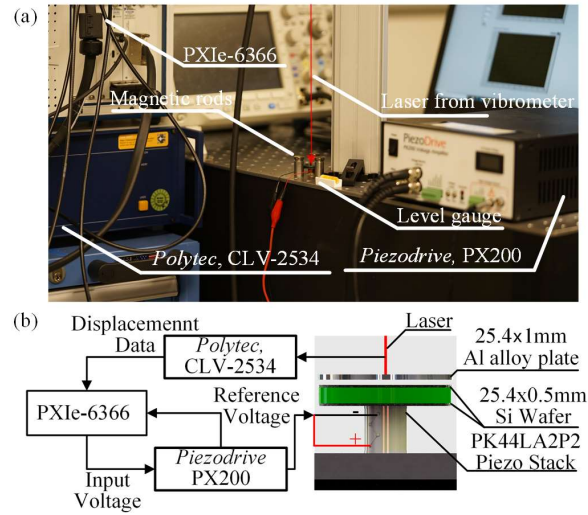

**FIG. 3.** (a) Experimental setup and (b) schematics of the actuator and measurement devices.

To evaluate the model accuracy, the predicted levitation height was first compared with experimental results to validate the dynamic stiffness. Since the dynamic force cannot be directly measured, the levitation height is used as comparison metrics by applying a known weight to the levitated object. The levitation height then can be analytically predicted by solving Eq. (7) as

$$\bar{h} = \sqrt{\frac{\pi R^2 p_a \Phi_f A_{mp}^2}{2mg}}. \tag{8}$$

The 1 mm thick aluminum plate weighed at 1.35 g, while the vibration frequency was set at 9 kHz. The predicated and measured levitation heights under different excitation amplitudes are compared in Fig. 4(a). The levitation phenomenon has been observed when the vibration amplitude exceeds 0.5 μm, which is specific to the load mass. As predicted, the levitation height shows almost a linear relationship with the excitation amplitude. As observed in our previous experiments, the levitation height solved directly by the Reynolds equations is often an overestimate since the levitation height has to be pre-given as a known condition to solve the differential equations for the pressure distribution. The time-averaged dynamic behavior of NFAL cannot be directly captured in the Reynolds equations, while the proposed model based on an equivalent nonlinear and asymmetric spring gives a more accurate prediction.

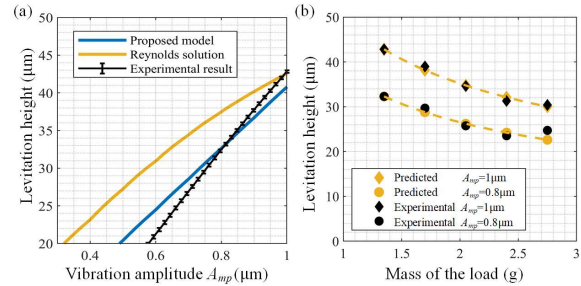

**FIG. 4.** Model verification results of predicted levitation height given different (a) excitation amplitudes and (b) levitated loads.

The second set of experiments measures the levitation height while changing the levitation mass, which can directly verify the effective stiffness $K$. For NFAL bearings, an important concern is that how the levitation height will change when the load is varied. By utilizing Eq. (8), where the levitation force is given as the prescribed load, the predicted levitation height can be obtained. In the experiments, we controlled the levitation force by adding mass in a quasi-static manner. The additional mass at the increment of 0.35 g was attached to the top side of the aluminum plate. The overall tested range was from 1.35 g to 2.75 g. The predicted and experimental results are compared in Fig. 4(b), which agree with each other. For a given excitation amplitude, the levitation height is inversely proportional to the square root of the load.

It is interesting to compare the results presented in Fig. 4(b) (compliance as the local slope) and the calculated effective stiffness presented in Figs. 2(c) and 2(d). First, the effective stiffness increases rapidly with the decrease in the levitation height ($1/\bar{h}^3$), as evidenced by the almost flat slope towards the right end of the curves in Fig. 4(b). Secondly, the effective stiffness, in theory, can be enhanced by either increasing the excitation vibration amplitudes or decreasing the levitation height as shown in Fig. 2(c). In a free levitation configuration, however, these values are coupled. Increasing the excitation amplitudes will also increase the levitation height if the load is unchanged, thus having a complex influence on the final stiffness.

Another major performance index is the sensitivity of NFAL bearings to the dynamic external disturbance. Particularly we experimentally investigate the levitation height variation due to the external displacement input at various frequencies. In this set of experiments, an additional sinusoidal modulation motion is added to the existing high-frequency excitation vibration of the bottom surface. The modulation can be regarded as a harmonical disturbance input, while the variation of the levitation height is measured. The frequency range of the modulation was set from 200 Hz to 2000 Hz. This frequency range is well below the dominant excitation frequency of 9 kHz to avoid aliasing and will not generate additional levitation force. Fig. 5(a) shows the measured vibration profiles when a modulation at 1200 Hz was added. The experiment conditions are summarized in Table 2.

Table 2 Experiment parameters for disturbance response analysis.

| Base vibration amplitude $A_{mp}$ (μm) | Excitation frequency $f$ (Hz) |
|---|---|
| 0.25, 0.375, 0.5 | 9000 |
| Modulation amplitude $X_d$ (μm) | Modulation frequency $f_d$ (Hz) |
| 1.0 | 200, 300, …, 2000 |

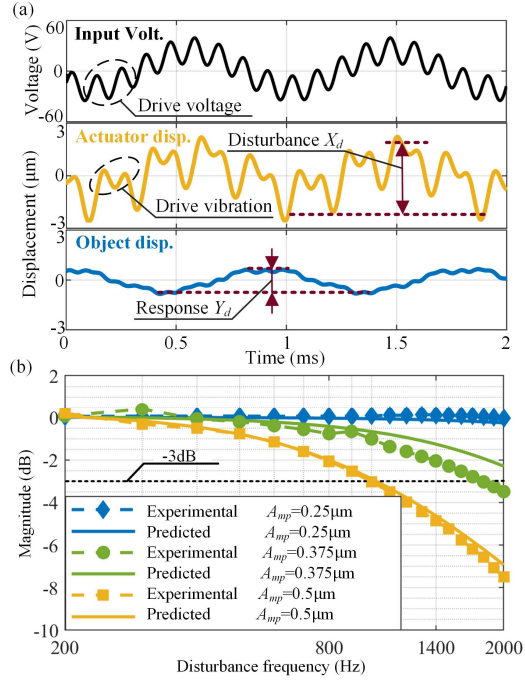

**FIG. 5.** (a) Measured levitation height variation due to the voltage modulation and (b) experimental and predicted frequency response of levitation height due to disturbance.

According to the transfer function of a 2$^{nd}$ order system, as presented in Fig. 1(a), the frequency response of the system (defined as the ratio of response to disturbance modulation amplitudes $Y_d/X_d$) can be derived as

$$T_{abs}(f_d) = \frac{Y_d}{X_d} = \left|\frac{\pi R^2 p_a \Phi_f \varepsilon^2 + j2\pi c f_d \bar{h}}{\pi R^2 p_a \Phi_f \varepsilon^2 - 4\pi^2 m f_d^2 \bar{h} + j2\pi c f_d \bar{h}}\right|, \qquad (9)$$

where the floating height $\bar{h}$ is determined by Eq. (8). The damping ratio $c$ is approximated following Ilssar and Bucher's work,[11] which is given by

$$c = \frac{3\pi R^4 \mu \varepsilon}{2\bar{h}^3}. \qquad (10)$$

The experimental and predicted results of the frequency response ($Y_d/X_d$) are compared in Fig. 5(b). From the results, for our particular setup, the system can be characterized as an over-damped second-order system, which shows no practical resonant frequency. The situation might be different if the levitated mass is much larger (with a presented resonant frequency for the air-mass system). The deviation of the predicted curves from the measured results is partly due to the simple assumption of the damping ratio, which might be a complex term with a velocity dependence. It is interesting to note that by increasing the excitation vibration amplitude, the system response actually attenuates faster due to the increased average levitation height (thus decreased dynamic and effective stiffness). When looking at the -3 dB threshold value, the worst-case scenario shows a 900 Hz bandwidth, which is quite impressive for industrial applications. This implies that if an impulse disturbance is applied, the system will return to the original balanced position much faster.

In summary, a dynamic model for NFAL is proposed in this study to define two system stiffness variables. The analytical solutions are derived for the two types of stiffness based on linearized solutions to the Reynolds equations. It can be concluded that:

(1) The dynamic stiffness $k$ is a reciprocal function of levitation height. It represents a non-linear and asymmetric spring, which hardens when squeezing the air while softens when releasing. The dynamic stiffness relates the excitation amplitude to dynamic force, but only at the excitation frequency for a very small squeeze ratio.

(2) The effective stiffness $K$ characterizes the relationship between the levitation force and average height, which is more closely related to the conventional definition of bearing stiffness. It is inversely proportional to the cubic of levitation height and relates to the high-frequency stiffness by the square of squeeze ratio.

(3) The modeled system demonstrates potential high stiffness at low levitation height, which will be comparable to commercial air bearings if scaled up with power and surface areas. It also demonstrates good bandwidth under dynamic disturbance, which is beneficial for industrial applications.

**Supplementary Material**

See supplementary material for the discussions on the physical meaning of frequency correction constant $\Phi_f$, detailed experimental setup, and additional experimental data.

**Acknowledgments**

This research was supported by the start-up fund from McCormick School of Engineering, Northwestern University, USA.

**Data availability**

The data that supports the findings of this study are available within the article and its supplementary material. Additional information is available from the corresponding author upon reasonable request.

*Supplementary Materials*
# Stiffness modeling for near field acoustic levitation bearings

Yaoke Wang[1] and Ping Guo[1,*]
*1Department of Mechanical Engineering, Northwestern University, Evanston, IL 60208, USA*
*\* Corresponding author: ping.guo@northwestern.edu*

### (1) Physical meaning of frequency correction term $\Phi_f$

We define a frequency correction term $\Phi_f$ in Eq. (5), which is a linear constant that relates the squeeze ratio to the normalized dynamic force. We further give a detailed explanation here on the physical meaning of this frequency correction term. By solving Eq. (3), the pressure distributions and evolutions $p(r,t)$ within the air gap are plotted in Fig. S1 for several represented conditions (all the simulation parameters are listed in Table 1).

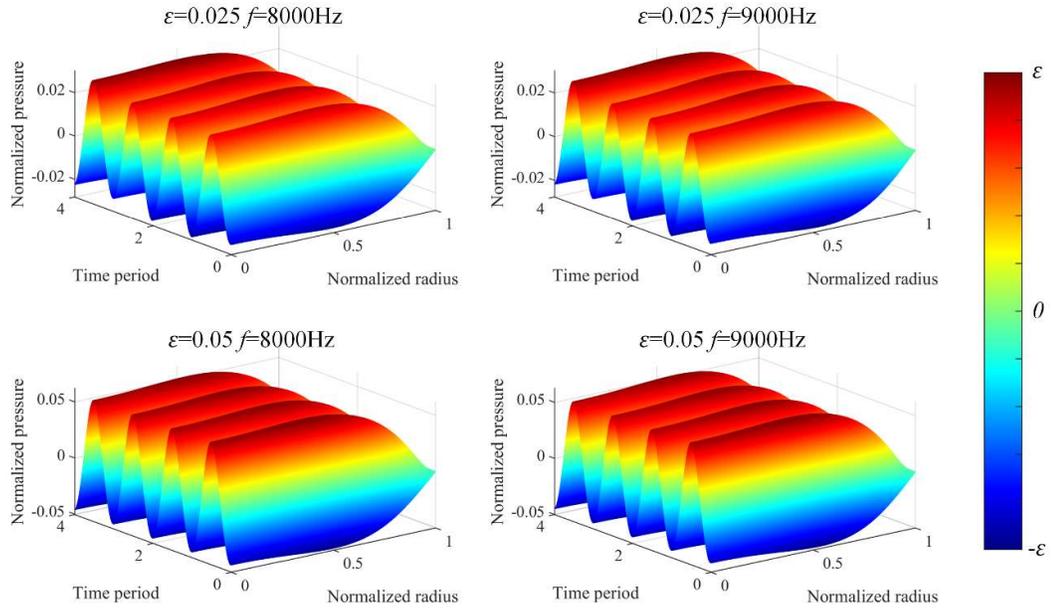

**FIG. S1** Simulated pressure distribution and evolution.

The different pressure distributions present a similar trend, where the normalized pressure at the center ($r=0$) is close to the squeezing ratio $\varepsilon$. Under the conditions of high vibration frequency, a critical point $R_m$ at which $\partial p/\partial r = 0$ emerges as a stagnation boundary that effectively reflects the airflow as shown in Fig. S2.

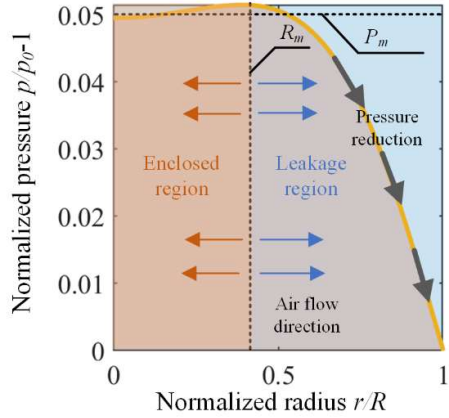

**FIG. S2** Schematic of the enclosed and leakage regions of air pressure distribution.

Since the average lateral air velocity is proportional to the pressure gradient, this critical point ($R_m$) separates the air domain into an enclosed region and a leakage region. Inside the enclosed region, the average pressure $p_m(t)$ can be approximated by the ideal gas equation following an isothermal process (Since the air gap is very thin, the volume change due to minor drift of $R_m$ within one period is negligible).

$$\frac{\partial (p_m + p_a)h}{\partial t} = 0, \tag{S1}$$

$$p_m(t) = \left(\frac{\bar{h}}{h} - 1\right)p_0 = \frac{p_a A_{mp} \sin(2\pi f t)}{h}. \tag{S2}$$

By utilizing $p_m$ as a reference pressure for the entire air film, the real-time levitation force and its dynamic amplitude can be approximated by introducing a frequency correction coefficient by

$$F(t) = \pi R^2 C p_m(t) = \frac{\Phi_f f \pi R^2 p_a}{h} A_{mp} \sin(2\pi f t), \tag{S3}$$

$$F_{amp} = \frac{1}{2}\Phi_f \pi R^2 p_a \left(\frac{A_{mp}}{\bar{h} - A_{mp}} + \frac{A_{mp}}{\bar{h} + A_{mp}}\right) \approx \Phi_f \pi R^2 p_a \varepsilon. \tag{S4}$$

Here, if $\Phi_f$ equals 1, it corresponds to a fully enclosed boundary condition where no air leakage is assumed. In a realistic condition, $\Phi_f$ is always smaller than one considering the existent of the leakage region. When the excitation frequency increases, the position of critical point $R_m$ expands, causing a decrease in the leakage region, or the increase in the value of $\Phi_f$.

In addition, the effects of object surface area on the frequency correction factor are further investigated. The calculated $\Phi_f$ is plotted in FIG. S3 for the different object radii of 12.7 mm (0.5"), 19.1 mm (0.75"), and 25.4 mm (1"). The frequency correction factor does depend on the object surface area, but will eventually converge to a constant at a higher excitation frequency. For a practical application of the proposed stiffness model, since most NFAL devices work at a fixed frequency at an ultrasonic frequency, this area/radius dependence will even be smaller. The frequency correction factor can either be calibrated using a few test cases or simply assumed to be a constant between 0.8 to 1. As shown in Fig. S3, the variation of $\Phi_f$ at 2 kHz due to the change of radius will be less than 10% for the three cases presented. This variation will further decrease with the increase of the surface area, due to the reduced influence from the boundary effects.

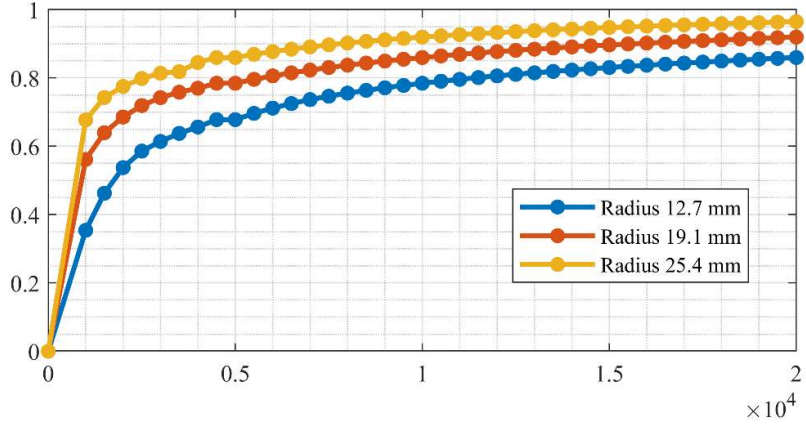

**FIG. S3** Calculated frequency correction term $\Phi_f$ for different object radii.

## (2) Additional discussions on the experimental setup

To eliminate unwanted external disturbance and resonant vibration, the whole setup sits on a vibration isolation table. The vibration of the actuation surface has been checked that the vibration amplitudes follow a linear relationship with the excitation voltage amplitude, as shown in Fig. S4. In addition, the vibration amplitudes at the center and edge locations are compared to be the same across the frequency range. These measures are to ensure no resonant vibration occurs within the test frequency range, so the air is squeezed by the pure translational motion as assumed in the simulation model. The frequency range (as well as the maximum excitation voltage amplitude) is limited by the current output capacity of the piezo amplifier (PX200, *Piezodrive*).

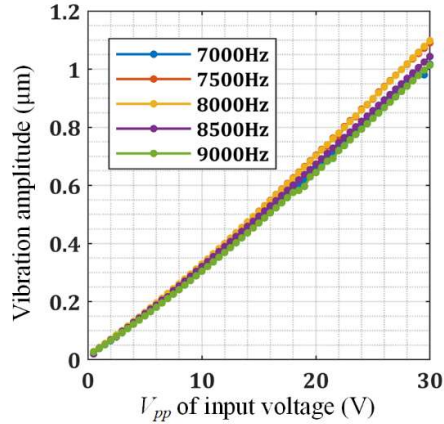

**FIG. S4** Measured vibration amplitudes of the actuation surface at different excitation voltages and frequencies.

## (3) Additional experimental data

Table. S1 Floating height at multiple frequencies and amplitudes

| Vibration amplitude (μm) | Frequency (Hz) | | | |
|---|---|---|---|---|
| | 6000 | 7000 | 8000 | 9000 |
| | Levitation height (μm) | | | |
| 0.25 | 6.56 | 5.45 | 4.39 | 2.31 |
| 0.27 | 7.40 | 6.25 | 5.23 | 3.16 |
| 0.28 | 8.21 | 7.07 | 6.08 | 4.05 |
| 0.30 | 9.00 | 7.82 | 6.94 | 4.93 |
| 0.32 | 9.82 | 8.64 | 7.78 | 5.81 |
| 0.34 | 10.63 | 9.45 | 8.64 | 6.69 |
| 0.35 | 11.44 | 10.28 | 9.51 | 7.59 |
| 0.37 | 12.24 | 11.04 | 10.41 | 8.49 |
| 0.39 | 13.01 | 11.88 | 11.28 | 9.39 |
| 0.40 | 13.77 | 12.67 | 12.19 | 10.33 |
| 0.42 | 14.55 | 13.48 | 13.06 | 11.25 |
| 0.44 | 15.28 | 14.30 | 13.97 | 12.19 |
| 0.45 | 16.03 | 15.07 | 14.88 | 13.13 |
| 0.47 | 16.75 | 15.86 | 15.80 | 14.10 |
| 0.49 | 17.46 | 16.65 | 16.69 | 15.06 |
| 0.51 | 18.14 | 17.43 | 17.59 | 16.03 |
| 0.52 | 18.81 | 18.22 | 18.50 | 17.00 |
| 0.54 | 19.48 | 18.98 | 19.39 | 17.97 |
| 0.56 | 20.12 | 19.75 | 20.29 | 18.93 |
| 0.57 | 20.76 | 20.52 | 21.18 | 19.88 |
| 0.59 | 21.40 | 21.27 | 22.08 | 20.89 |
| 0.61 | 22.02 | 22.02 | 22.98 | 21.84 |
| 0.63 | 22.59 | 22.77 | 23.87 | 22.79 |
| 0.64 | 22.96 | 23.50 | 24.74 | 23.75 |
| 0.66 | 23.51 | 24.19 | 25.62 | 24.72 |
| 0.68 | 24.05 | 24.89 | 26.48 | 25.68 |
| 0.69 | 24.64 | 25.54 | 27.33 | 26.63 |
| 0.71 | 25.11 | 26.25 | 28.19 | 27.58 |
| 0.73 | 25.60 | 26.93 | 29.03 | 28.54 |
| 0.74 | 26.12 | 27.58 | 29.88 | 29.47 |
| 0.76 | 26.67 | 28.22 | 30.63 | 30.42 |
| 0.78 | 27.12 | 28.85 | 31.47 | 31.35 |
| 0.80 | 27.62 | 29.54 | 32.21 | 32.29 |
| 0.81 | 28.06 | 30.15 | 33.05 | 33.19 |
| 0.83 | 28.51 | 30.75 | 33.83 | 34.11 |
| 0.85 | 28.98 | 31.34 | 34.55 | 35.02 |
| 0.86 | 29.43 | 31.92 | 35.31 | 35.91 |
| 0.88 | 29.86 | 32.49 | 36.05 | 36.80 |
| 0.90 | 30.33 | 33.04 | 36.78 | 37.68 |
| 0.91 | 30.72 | 33.58 | 37.49 | 38.56 |
| 0.93 | 31.12 | 34.14 | 38.17 | 39.42 |
| 0.95 | 31.54 | 34.68 | 38.85 | 40.28 |
| 0.97 | 32.03 | 35.20 | 39.54 | 41.15 |
| 0.98 | 32.43 | 35.73 | 40.17 | 41.97 |
| 1.00 | 32.91 | 36.21 | 40.80 | 42.83 |

Table. S2 Floating height with changing mass at 9000 Hz.

| Vibration amplitude (μm) | Mass (g) | | | | |
|---|---|---|---|---|---|
| | 1.35 | 1.70 | 2.05 | 2.40 | 2.75 |
| | Floating height (μm) | | | | |
| 1.00 | 46.20 | 39.99 | 34.62 | 32.18 | 30.40 |
| 0.98 | 45.32 | 39.18 | 33.84 | 31.52 | 30.30 |
| 0.97 | 44.49 | 38.36 | 33.06 | 30.87 | 30.07 |
| 0.95 | 43.60 | 37.52 | 32.25 | 30.15 | 29.72 |
| 0.93 | 42.72 | 36.69 | 31.46 | 29.45 | 29.31 |
| 0.91 | 41.85 | 35.86 | 30.68 | 28.74 | 28.72 |
| 0.90 | 40.98 | 35.03 | 29.90 | 28.00 | 28.17 |
| 0.88 | 40.08 | 34.19 | 29.07 | 27.33 | 27.55 |
| 0.86 | 39.17 | 33.32 | 28.24 | 26.62 | 26.88 |
| 0.85 | 38.26 | 32.43 | 27.43 | 25.87 | 26.06 |
| 0.83 | 37.33 | 31.56 | 26.57 | 25.15 | 25.44 |
| 0.81 | 36.40 | 30.68 | 25.75 | 24.41 | 24.72 |
| 0.80 | 35.46 | 29.77 | 24.88 | 23.66 | 24.08 |

Table. S3 Floating height with changing mass at 8000 Hz.

| Vibration amplitude (μm) | Mass (g) | | | | |
|---|---|---|---|---|---|
| | 1.35 | 1.70 | 2.05 | 2.40 | 2.75 |
| | Floating height (μm) | | | | |
| 1.00 | 40.93 | 36.04 | 31.26 | 29.39 | 28.19 |
| 0.98 | 40.30 | 35.42 | 30.65 | 28.84 | 28.08 |
| 0.97 | 39.67 | 34.75 | 30.03 | 28.27 | 27.79 |
| 0.95 | 38.98 | 34.11 | 29.39 | 27.64 | 27.30 |
| 0.93 | 38.30 | 33.45 | 28.73 | 27.04 | 26.71 |
| 0.91 | 37.62 | 32.77 | 28.05 | 26.39 | 26.12 |
| 0.90 | 36.90 | 32.01 | 27.37 | 25.75 | 25.53 |
| 0.88 | 36.17 | 31.29 | 26.67 | 25.09 | 24.97 |
| 0.86 | 35.44 | 30.56 | 25.97 | 24.40 | 24.35 |
| 0.85 | 34.68 | 29.82 | 25.26 | 23.75 | 23.84 |
| 0.83 | 33.96 | 29.06 | 24.53 | 23.06 | 23.10 |
| 0.81 | 33.18 | 28.30 | 23.79 | 22.39 | 22.53 |
| 0.80 | 32.33 | 27.54 | 23.06 | 21.72 | 21.95 |

Table. S4 Frequency response of disturbance $Y_d$ (the base excitation frequency is at 9000 Hz).

| Modulation frequency (Hz) | Modulation amplitude (μm) (input) | | |
|---|---|---|---|
| | 0.25 | 0.375 | 0.5 |
| | Upper surface modulation amplitude (μm) (output) | | |
| 2000 | 0.89 | 0.61 | 0.46 |
| 1900 | 0.89 | 0.63 | 0.47 |
| 1800 | 0.90 | 0.65 | 0.49 |
| 1700 | 0.90 | 0.67 | 0.50 |
| 1600 | 0.90 | 0.69 | 0.52 |
| 1500 | 0.90 | 0.70 | 0.54 |
| 1400 | 0.90 | 0.73 | 0.57 |
| 1300 | 0.90 | 0.75 | 0.60 |
| 1200 | 0.90 | 0.77 | 0.62 |
| 1100 | 0.90 | 0.79 | 0.66 |
| 1000 | 0.90 | 0.81 | 0.69 |
| 900 | 0.90 | 0.83 | 0.73 |
| 800 | 0.90 | 0.85 | 0.77 |
| 700 | 0.91 | 0.87 | 0.81 |
| 600 | 0.91 | 0.89 | 0.86 |
| 500 | 0.91 | 0.91 | 0.90 |
| 400 | 0.91 | 0.93 | 0.94 |
| 300 | 0.92 | 0.94 | 0.96 |
| 200 | 0.92 | 0.96 | 1.04 |